\documentclass[12pt]{article}

\usepackage{amssymb}
\usepackage{amsmath}
\usepackage{amscd}
\usepackage{latexsym}
\usepackage{graphicx}

\usepackage{cite}
\usepackage{enumitem}

\newcommand{\be}{\begin{equation}}
\newcommand{\ee}{\end{equation}}

\newcommand{\dlt}{\delta}
\newcommand{\prt}{\partial}
\newcommand{\br}{{\bf r}}
\newcommand{\dgr}{\dagger}

\newcommand{\vp}{\varphi}
\newcommand{\ep}{\varepsilon}

\newcommand{\ra}{\rightarrow}
\newcommand{\cH}{{\cal H}}

\newcommand{\rgl}{\rangle}
\newcommand{\lgl}{\langle}

\begin{document}

\begin{center}

{\Large{\bf Saga of Superfluid Solids} \\ [5mm]

Vyacheslav I. Yukalov$^{1,2,*}$ } \\ [3mm]

$^1${\it Bogolubov Laboratory of Theoretical Physics, \\
Joint Institute for Nuclear Research, Dubna 141980, Russia \\ [2mm]

$^2$Instituto de Fisica de S\~ao Carlos, Universidade de S\~ao Paulo, \\
CP 369, S\~ao Carlos 13560-970, S\~ao Paulo, Brazil   \\  [2mm]

$^*${\bf Author e-mail}: yukalov@theor.jinr.ru }

\end{center}

\vskip 2cm

\begin{abstract}
The article presents the state-of-the-art and reviews the literature on 
the long-standing problem of the possibility for a sample to be at the same time 
solid and superfluid. Theoretical models, numerical simulations, and experimental 
results are discussed.
\end{abstract}

\vskip 1cm

{\parindent = 0pt
{\bf Keywords}: solid state; superfluid state; diagonal order; off-diagonal order;
translational symmetry; global gauge symmetry }

\vskip 3cm

\section{Introduction}

The term "superfluidity" was coined by Kapitza, who actually discovered the phenomenon 
of superfluidity in liquid Helium, in his paper \cite{Kapitza_165} followed by the paper 
by Allen and Misener \cite{Allen_166}. Kapitza had set up an experiment with two cylinders 
that were connected by a thin tube. Only below the $\lambda$ point, helium was flowing 
easily through the tube, suggesting a strikingly low viscosity. The first explanation of 
superfluidity, in the frame of the two-fluid model, was suggested by Landau 
\cite{Landau_167,Landau_168,Landau_169} and Tisza \cite{Tisza_170}. The key evidence 
establishing the two fluid model of superfluidity was given by Andronikashvili
\cite{Andronikashvili_171,Andronikashvili_172,Andronikashvili_173} 
studying the period and damping of torsional oscillations of stacked closely spaced 
rotating disks. More details about the interesting history concerning the discovery of
and first works on superfluidity can be found in Refs. 
\cite{Lifshitz_174,Andronikashvili_175,Balibar_176,Schmitt_177}.  

However, the subject of this review is not the description of superfluidity in liquids, 
but a rather unexpected problem considering the possibility of superfluid effects in 
solids. This review gives a brief account on the history and the present day situation 
on the possibility for a solid to be at the same time superfluid. This puzzle has a long 
history that at the present time is intensively explored. Historically, the assumption 
that a solid could display superfluid properties was considered for solid helium, and 
the majority of works have been  devoted to this material. Recently, the simultaneous 
breaking of translational symmetry, typical of crystals, and global gauge symmetry, 
usually accompanying superfluids, has been displayed for trapped dipolar gases under 
special conditions (Bose-Einstein condensation, tuned atomic interactions, and aligned 
magnetic dipoles). 

The review attempts to present the overall portrait, touching only the basic ideas, 
sufficient for grasping the outline of the whole picture, without much plunging into 
specific details. The brevity of the narration is compensated by a quite comprehensive 
list of literature. Some more information on concrete details from the works before 
2012 can be found in the review articles \cite{Prokofiev_1,Kuklov_178,Boninsegni_2}. 

The main aim of the present review is to suggest a general description that, although 
being precisely formulated, would be understandable for wide audience, where not all 
are professionals in the field of quantum condensed matter.

\section{What is a superfluid solid}

To the first glance, the topic announced in the title sounds as an oxymoron. The common
wisdom teaches us that a solid is rigid, keeping a fixed shape, while a superfluid is a 
substance having no fixed shape and no viscosity, easily flowing without friction. Thus 
in the Merriam-Webster dictionary a solid is defined  as "{\it a substance that does not 
flow perceptibly under moderate stress, has a definite capacity for resisting forces 
(such as compression or tension) which tend to deform it, and under ordinary conditions 
retains a definite size and shape}". In the same dictionary a superfluid is classified 
as "{\it an unusual state of matter...characterized by apparently frictionless flow}". 

Of course the ability to flow without friction is not the sole characteristic of a
superfluid which possesses other exotic properties. Thus an important feature of a 
superfluid is the appearance of quantized vortices under rotation. A quantum vortex 
differs from the familiar classical vortices in turbulent liquids, such as water, by the 
quantization of circulation, when the integral of velocity around the core of a vortex 
is proportional to $(2 \pi \hbar/m) n$, where $n$ is an integer. The other remarkable 
property of a superfluid is the fountain effect that causes liquid helium, being heated,  
to flow up the sides of open containers. 
  
With regard to friction, it is useful to mention that there exist classical liquids 
composed of self-propelled particles that can organize themselves to counterbalance the 
energy loss resulting from viscous dissipation and thereby dramatically lower the fluid's 
viscosity, driving it to vanish or even to become negative \cite{Marchetti_179}. However, 
a frictionless flow with no energy dissipation is usually considered as a hallmark of 
superfluidity (and superconductivity). 

It looks that such contradictory terms as solid and superfluid could not be met in 
conjunction. Nevertheless, the idea that solids could have some superfluid properties 
has been advanced in theoretical works and has been intensively studied both 
theoretically and experimentally till nowadays. 

To correctly understand the subject, it is necessary, first of all, to recollect the 
definitions of what is {\it solid state} and what is {\it superfluid state}. 

In the common parlance, solid state is a rigid state of matter, being able to keep its 
shape for very long time, provided that external forces do not overpass threshold values. 
Such external forces, that could destroy a solid sample, could be pressure, temperature, 
or chemical impact. Otherwise, when these external forces are below their thresholds, a 
solid sample can survive for very long time. 

Usually, solids are arranged in crystalline structures. Such spatially periodic or 
quasiperiodic structures are said to possess diagonal long-range order. One also says 
that in crystals translational uniform symmetry is spontaneously broken, as compared 
with translationally uniform liquids. It is important that the symmetry breaking is 
spontaneous, that is, self-organized. Thus, if one imposes an external periodic field 
upon liquid or gas, they acquire some periodic properties. But anyway, they are liquid 
or gas, not solids.  

Periodic structures can sometimes self-organize in nonequilibrium liquid or gaseous 
systems for a short period of time. Clearly, such temporally periodic systems have 
nothing to do with solids, even if their periodicity has appeared spontaneously. For 
example, quantum Bose gases can be Bose-condensed, with exhibiting superfluid properties. 
Upon their interactions, during some nonequilibrium processes, they can form temporal 
interference patterns. In that sense the system demonstrates spontaneous breaking of 
uniform translational symmetry. Such gases can be temporarily superfluid and periodically 
arranged in an interference pattern. However, they are not solids, they are gases. And 
they are nonequilibrium, exhibiting interference patterns for a finite, usually very 
short, time. 

There exist as well the terms "amorphous solids" or "glasses" that do not exhibit 
long-range periodic spatial order. They may look like genuine solids, being often rather 
rigid and being able to keep their shape for long time. However these materials are 
not absolutely stable, but metastable, and tend to crystallize with time, although 
this time can be very long, sometimes like tens or hundreds years.            

The intermediate case between the crystals, with ideally periodic lattices, and 
amorphous solids, with random locations of their constituents, are the crystals with 
defects. Crystals with defects, such as vacancies, interstitials, and dislocations, 
are mainly periodic, but the strict periodicity is absent around defects. Actually 
the majority of real crystals contain some defects, except maybe rare specially 
prepared samples.    

Summarizing, it is possible to accept the following convention: {\it Solid state is 
a rigid state of matter that in equilibrium is stable or metastable in a finite region 
of parameters, such as pressure and temperature, or some external fields, where it is 
able to keep its shape for long time}.
  
Contrary to solids, {\it superfluids are fluids that cannot keep a rigid shape, but 
can flow without friction and exhibit such features as quantized vortices and fountain 
effect}. The widely known examples are $^3{\rm He}$ and $^4{\rm He}$.

A solid that would combine the features of rigidity and superfluidity would deserve the 
name of {\it superfluid solid}. In literature, one often calls such superfluid solids as 
"supersolids", as suggested by Matsuda and Tsuneto \cite{Matsuda_180} and Mullin 
\cite{Mullin_181}. This name, however, looks a bit confusing. The standard meaning of 
the word "super" accentuates the given property, but does not contradict it. For instance, 
"superradiance" means superstrong radiance. "Superconductivity" implies superstrong 
conductivity. "Superfluidity" signifies superstrong fluidity. Then "supersolidity" should 
assume superrigid solidity. While, vice versa, one talks about a solid with some 
superfluid properties. Also, the term "supersolid" has already been used for many years 
with respect to crystals in space dimensionality larger than three \cite{Oxford_18}.       
 
One often names a "supersolid" any system, where simultaneously there occurs uniform 
translational symmetry breaking, typical of solids, and global gauge symmetry breaking,
typical of superfluids. In that sense, interfering Bose-condensed gases should be termed 
"supersolids", which sounds a bit strange. Interfering gases are anyway gases, even if 
they are spatially periodic. In addition, superfluidity does not necessarily require
global gauge symmetry breaking.     

Thus, a superfluid solid can be classified as such, provided it enjoys the following 
mutually complimenting properties.

\begin{enumerate}[label=(\roman*)]
\item
It has to be a solid, that is a rigid system being able to keep for long time its shape 
in a finite range of external conditions. Long time implies the time that is much longer 
than any of the characteristic internal times, such as interaction time and local 
equilibration time. 

\item
It has to possess the above properties being in an equilibrium state, meaning either 
an absolutely equilibrium state or a metastable state. This does not impose the requirement 
of necessarily being periodic. Amorphous solids also are solids. And crystals with defects
are not ideally periodic.  

\item
It has to demonstrate, inside its volume, superfluidity as a frictionless motion of 
matter. This maybe connected with the spontaneous breaking of gauge symmetry, at least  
locally, although this is not compulsory.      
\end{enumerate}

These properties are kept in mind throughout the review when we talk about superfluid 
solids. To stress it once more, the latter are not necessarily ideally periodic, but have 
to be rigid and stable. Interfering periodically modulated Bose gases are interesting 
objects that, however, are not solids. In addition, they are usually unstable. 
Superfluidity occurring in external periodic potentials, for instance in optical lattices,
also does not constitute superfluid solids, since such externally imposed lattices are not 
self-organized.

\section{Model of coherent crystal}

A Bose system at zero temperature and weak interactions can be almost completely Bose 
condensed, which means that the system is in a coherent state. Then such a system can be
described in the quasiclassical approximation. The standard Hamiltonian in this 
approximation reads as
$$
H = \int \vp^*(\br,t) \left( - \; \frac{\nabla^2}{2m} + U \right) \vp(\br,t) \; d\br \; +
$$
\be
\label{1}
+ \; 
\frac{1}{2} \int \vp^*(\br,t) \vp^*(\br',t) \Phi(\br-\br') \vp(\br',t) \vp(\br,t) \; 
d\br d\br' \; ,
\ee
where $\varphi ({\bf r},t)$ is a {\it complex-valued function} (not operator), called 
condensate function, $U = U({\bf r}, t)$ is an external potential, and $\Phi({\bf r})$ 
is an interaction potential. The Planck constant is set to one. The equation of motion 
\be
\label{2}
i\; \frac{\prt}{\prt t}\; \vp(\br,t) = \frac{\dlt H}{\dlt \vp^*(\br,t)}
\ee
yields the Nonlinear Schr\"{o}dinger equation
\be
\label{3}
 i\; \frac{\prt}{\prt t}\; \vp(\br,t) =   
\left( - \; \frac{\nabla^2}{2m} + U \right) \vp(\br,t) + 
\int  \Phi(\br-\br') \; | \vp(\br',t) |^2 d\br' \vp(\br,t) \; .
\ee
Equation (\ref{3}) was introduced by Bogolubov in his well known book {\it Lectures 
on Quantum Statistics} \cite{Bogolubov_49}, which has been republished many times, 
for instance in \cite{Bogolubov_67,Bogolubov_70,Bogolubov_15}. Gross 
\cite{Gross_55,Gross_57,Gross_58,Gross_60,Gross_60a,Gross_61,Gross_63} has intensively 
studied this equation showing that it possesses nonuniform solutions, including stationary 
periodic solutions \cite{Gross_57,Gross_58,Gross_60,Gross_60a}, when particle interactions 
contain an attractive part. Since the coherent condensate function corresponds to the 
Bose-Einstein condensate that supports superfluidity, the periodic Gross solution describes 
a periodic superfluid system. This could be treated as the first model of superfluid solid, 
provided it would be proved that it is really related to a solid, but not just to a periodic 
gas. However, the found periodic solution looks to be more appropriate to a periodically 
modulated superfluid than to a solid \cite{Gross_57,Gross_58,Gross_60}.

Similar idea of coherent crystallization was considered later for more refined models  
\cite{Kirzhnits_71,Vozyakov_77}. The essential point in such models is the requirement that
particles, being in the same coherent state, could freely move either over the whole volume 
of the system, or at least over some of its macroscopic parts. Without such a motion, the
system cannot be coherent and also Bose-Einstein condensation cannot occur \cite{Penrose_56}.

\section{Model of vacancion superfluidity}

Andreev and Lifshits \cite{Andreev_69} suggested that superfluidity can arise in quantum 
crystals with defects. Since vacancies are essentially more mobile than impurities
\cite{Pushkarov_91}, they can form a kind of a quantum liquid of vacancions inside the 
crystal and superfluidity can develop \cite{Andreev_76,Andreev_82}. 

Chester \cite{Chester_69,Chester_70} argued that Bose-Einstein condensation with zero 
momentum (hence superfluidity) should appear in any quantum crystal described by the 
ground-state many-body wave function of the Jastrow type
\be
\label{4}
 \Psi(\br_1,\br_2,\ldots) \; \propto \; \prod_{i\neq j} f(r_{ij} ) \;  ,
\ee
in which $r_{ij} \equiv |{\bf r}_i - {\bf r}_j|$ and 
$$
f(r) = \exp\left\{ -\; \frac{1}{2}\; u(r) -\chi(r) \right\} \;   ,
$$
where $u(r)$ is real, bounded from below and has a finite range, that is to say
$$
u(r) \; \propto \; \frac{const}{r^{3+\ep}} \qquad ( r \ra \infty) \;   ,
$$
with $\varepsilon > 0$, while $\chi (r)$ is finite and, either is zero or, for large $r$, 
behaves as
$$
\chi(r) \; \propto \; \frac{const}{r^2} \qquad ( r \ra \infty) \;    .
$$
But in an ideal crystal, according to Penrose \cite{Penrose_56}, Bose-Einstein condensation 
cannot arise. 

A number of experimental attempts has been undertaken trying to discover superfluidity in
solid $^4$He. An overview of these unsuccessful searches, accomplished till 1992, is given 
by Meisel \cite{Meisel_92}.

\section{Nonclassical rotation inertia}

Leggett \cite{Leggett_70} mentioned that the tentative appearance of superfluid fraction
in a solid could be found by measuring the moment of inertia $I(T)$ of the studied system 
as a function of temperature, as has been suggested earlier for liquids by Andronikashvili 
\cite{Andronikashvili_171,Andronikashvili_172,Andronikashvili_173}. Thus the rotational 
moment of inertia can be connected with the superfluid fraction by the relation
\be
\label{5}
 I(t) = I_0 \left[ 1 \; - \; \frac{\rho_s(T)}{\rho} \right ] \;  ,
\ee
where $I_0(T)$ is the classical moment of inertia for the whole system and $\rho$ is average
density. Upon the appearance of superfluid density $\rho_s(T)$, the rotational moment of 
inertia decreases and, respectively, the oscillation frequency of the sample, proportional 
to $1/\sqrt{I(T)}$, increases, while the resonant oscillation period 
$T_{osc} = 2 \pi /\omega \propto \sqrt{I(T)}$ decreases. 

Kim and Chan \cite{Kim_04,Kim_04a,Kim_05,Kim_06} accomplished a series of experiments with 
a torsional oscillator technique for $^4$He, claiming the discovery of superfluidity at low 
temperatures below about $200$ mK. These measurements were followed by many other experiments
announcing their agreement with the Kim-Chan results \cite{Rittner_06,Rittner_07,Aoki_07, 
Clark_07,Kondo_07,Clark_08a,Rittner_08,Mulders_08,West_09,Kitamura_11}. Similar nonclassical 
response was found in para-hydrogen clusters \cite{Clark_06,Mezzacapo_08,Li_10}.

However, recent more accurate experiments with $^4$He have found no measurable period drop 
that could be attributed to nonclassical rotational inertia. The authors have come to 
the conclusion that the rotational period drop observed in previous torsional oscillator 
studies was actually measuring the temperature dependence of structural effects, such as 
the shear modulus stiffening \cite{Kim_12,Kim_14,Choi_15,Fear_16}, and not superfluid 
effects. As will be explained in better detail below, the main story appears to rely on 
dislocations and flux along dislocation cores - effects apparently too small to be visible 
in rotational experiments.

\section{Shear modulus stiffening}

It was experimentally found that below about $200$ K the shear modulus of solid $^4$He 
increases with lowering temperature, strongly resembling the temperature behavior of the 
oscillation frequency. The increase in the shear modulus is likely due to the stiffening 
of the dislocation network via immobilization of dislocations on defects, such as admixture 
of $^3$He. The increase can be rather drastic, up to $20 \%$. This increase in the shear 
modulus mimics the increase of the frequency \cite{Day_07,Corboz_08,Clark_2008,Day_09,
Day_10,Syshchenko_10,Reppy_10,Eyal_10,Su_10,Pratt_11,Aleinikava_12,Mi_12,Zhou_12,Reppy_12,
Fefferman_12,Eyal_16,Tsuiki_18,Choi_18}. Thermal conductivity in polycrystalline hcp $^4$He 
also increases \cite{Zneev_11}.

These results are in agreement with the study by Bishop et al. \cite{Bishop_182} 
who measured the moment of inertia of hcp $^4$He crystals from $25$ mK to $2$ K. With a
precision of five parts in $10^6$ they found no evidence for a nonclassical rotational 
inertia. This indicates that if a superfluid solid exists, it has a fraction less than 
$5 \times 10^{-6}$. Thus the increase of the oscillator frequency at low temperatures 
in solid $^4$He does not signify the arising superfluidity, but is caused by the changes 
in the shear modulus.  

The plastic properties of hcp $^4$He are different at low temperatures below $0.5$ K and 
above this temperature \cite{Cheng_183}. Plasticity involves the motion and multiplication 
of dislocations. At low temperatures the solid responds elastically only for strains up 
to approximately $\epsilon < 0.2 \%$. Sudden stress drops emerge at higher strains, 
accompanied by the acoustic emission expected from dislocation avalanches. The dimension 
of the slip regions varies over at least $3$ orders of magnitude, up to several millimeters. 
The mobile dislocations travel at speeds close to the sound speed, much faster than the 
damped dislocations in conventional solids. The low-temperature dislocation avalanches can
increase the helium's elastic shear modulus. The avalanches disappear above $0.5$ K, 
replaced by smooth creep deformation involving thermally excited motion of dislocations. 
A detailed description of experiments on plastic deformation and flow in solid $^4$He and 
$^3$He are described by Beamish \cite{Beamish_184}.

\section{Important role of disorder}

In the continuing search for possible occurrence of superfluid solids, one usually connects 
the possible existence of such a matter with the necessity for it to represent in some sense 
a disordered crystalline structure. It seems that disorder of a crystalline lattice plays a 
crucial role in the possible appearance of superfluidity in a solid. Theoretical papers 
claiming that superfluid solids could be perfect crystals 
\cite{Saslow_05,Galli_RR_05,Saslow_J_06,Ye_06,Josserand_PR_07,During_JPR_12},
are usually either based on phenomenological models or use approximate methods allowing for 
the existence of Bose-Einstein condensate. Among the latter methods, it is possible to 
mention variational approaches based on Jastrow wave functions or shadow wave functions, 
which do demonstrate the presence of Bose condensate. 

It is now widely accepted that, if superfluid effects are expected to arise in a solid, 
this compulsorily requires the existence of some kind of disorder \cite{Balibar_08}. For 
instance, this could be quantum vacancies or interstitials, as it has been assumed by 
Andreev and Lifshits \cite{Andreev_69} and others 
\cite{Pokofev_S_05,Galli_R_05,Rossi_VGR_08,Pessoa_KV_09,Toda_GKKOS_10}, 
dislocations \cite{Rossi_VGR_10}, or grain boundaries and surfaces 
\cite{Dash_W_05,Penzev_YK_08,Gaudio_CRP_08,Cappeluti_RGP_08}. This can be a nanoporous 
glass, where a local Bose-Einstein condensation in nanopores is observed
\cite{Shirahama_YS_08,Yamamoto_SS_08,Yamamoto_SS_08a}, or an amorphous solid with an 
absolutely irregular structure, that is a kind of a glass 
\cite{Andreev_07,Grigorev_MRRRRST_ 07,Graf_BNGT_09,Hunt_PGYBD_09,Su_GB_10,Birshenko_MRV_12,
Andreev_12}.  

It is understood that a solid could exhibit superfluid properties only in the presence
of some kind of local disorder.

\section{Results of scattering experiments}

Solid helium, with both hcp and bcc structures, has been thoroughly investigated by 
scattering experiments in the range below $200$ mK in the attempts to find indications 
of some changes related to alleged superfluidity in solids. Neutron diffraction 
experiments, measuring the mean-square atomic displacement in hcp phase solid, found 
no evidence for an anomaly \cite{Blackburn_GSHBDFKD_07,Kirichek_09}.    

Neutron scattering experiments, measuring atomic momentum distribution $n(k)$, showed no 
sign of Bose-Einstein condensation in solid helium \cite{Diallo_PAKTG_07,Diallo_AKTG_09}. 
Neutron scattering experiments have also been used for studying the dynamic structure 
factor of amorphous solid helium. No liquid-like phonon-roton modes or other sharply 
defined modes at low energy or modes unique to a quantum amorphous solid that might 
suggest superflow are observed \cite{Bossy_OSG_12}. The maximal studied mode energy was 
lower than $1$ meV, that is $\omega(k) < 10$ K.

X-ray synchrotron radiation was used for measuring peak intensities and lattice parameters 
of solid $^4$He, showing no indication of superfluid effects \cite{Burns_MLCSKP_08}.  

However, there were reports observing, by means of inelastic neutron scattering, structural 
fluctuations in bcc solid helium \cite{Polturak_MBF_03,Pelleg_SLP_06} and coherent 
delocalized roton-like modes in hcp solid helium associated with delocalized atoms 
\cite{Blackburn_1}. See also discussion in review \cite{Galli_2}.

\section{Mass flux experiments}

If a solid sample would enjoy superfluid properties, it should exhibit mass transport upon 
applying a pressure gradient to the sample. This idea was checked in experiments 
\cite{Day_3} with solid He at low temperature below $200$ mK, where superfluid 
properties could be expected. These experiments showed no indication of superfluid flow 
in response to a pressure difference. The same conclusion of the absence of pressure-driven  
superfluid flow was obtained in low-frequency experiments \cite{Rittner_4}.

A different setup was used in the experiments of mass flux measurements 
\cite{Ray_5,Ray_6,Ray_7,Vekhov_8,Vekhov_9,Vekhov_10,Vekhov_11,Hallock_12} summarized by
Hallock \cite{Hallock_185}. The idea relies on the fact that superfluid helium in a 
micro-porous environment at a given temperature freezes at an elevated pressure
compared with the freezing pressure for bulk $^4$He. The micro-porous material Vycor,
filled by superfluid, has a very low thermal conductivity which allows the reservoirs of 
superfluid at temperatures near $1.5$K to be available to allow application of pressure 
and temperature differences between the reservoirs on the two sides of the apparatus, 
while a much colder region of solid $^4$He is in between. That is, it is possible to 
create an environment in which bulk solid helium, at pressures above the melting curve, 
can be in direct contact with superfluid Helium inside the porous material Vycor. Two 
Vycor rods containing superfluid Helium were inserted at the ends of a solid sample of 
Helium filling the horizontal cylindrical region of the cell. The basic question to be 
answered was whether atoms from one superfluid reservoir would pass into solid helium 
and as a result cause atoms to enter the other reservoir.  

This setup, called superfluid-solid-superfluid, allows injection of $^4$He atoms from 
superfluid directly into the solid hcp $^4$He. Manipulating mass injection and temperature, 
it is possible to create the chemical potential difference between the sample ends and 
produce a kind of superfluid flow. Samples grown at lower pressures show flow that 
depends on sample history, while samples grown at higher pressures show no clear evidence 
for any such flow. The flow was observed at temperatures below $550$ mK and at pressures 
below approximately $27$ bar. Above $600$ mK, the flux becomes too small to be measured. 
It falls abruptly in the vicinity of a blocking temperature $T_d$ depending on the 
concentration of the $^3$He admixture blocking the flux. At low enough concentration the 
flux rises with lower temperature after the fall, but at high enough concentration of 
the $^3$He admixture the flux becomes extinguished. 

The absolute value of the measured mass flux, in addition to temperature and applied 
pressure, depends on the specific sample and its history, but the general temperature 
dependence is similar for most of the samples studied. A partial anneal of the sample 
typically causes a reduction in the flux. The magnitude of the flux is measured in 
units g per mm$^2$ s. Thus in experiments of Halock et al. \cite{Hallock_185} the 
maximal flux was $1.5\times 10^{-9}$ g/mm$^2$ s. A much higher flux of approximately
$7.2\times 10^{-8}$ g/mm$^2$ s was seen for the sample with $8\mu$ thickness 
\cite{Shin_186}. However in the latter experiment the mass flow extinction was absent. 
In the work \cite{Shin_187} with solid samples of about $2$ mm thickness, a typical flux 
of approximately $2.5\times 10^{-8}$ g/mm$^2$ s is found. In general, it has been shown 
that thinner samples result in higher values of the flux, with the mass flow rate 
decreasing logarithmically with the thickness of the solid $^4$He. 

Similar results were obtained in other setups. Thus the work \cite{Cheng_188} reports the 
results of flow experiments in which two chambers containing solid $^4$He are connected 
by a superfluid Vycor channel, which is called solid-superfluid-solid junction. In the 
other experiment \cite{Cheng_189} the authors observed a flow in the setup where Vycor
has been eliminated, allowing to study the intrinsic flow in solid $^4$He without the 
complications introduced by the presence of superfluid and the associated solid-liquid 
interfaces. Applying a pressure gradient at one side, the flow was measured at the other 
side of the sample. As in previous experiments, the flow appeared below $600$ mK and 
increased with decreasing temperature before it was suppressed at a low temperature $T_d$
depending on the concentration of $^3$He admixture. These measurements show that mass flow 
in solid helium does not require superfluid leads but can be generated directly by pressure 
differences created by mechanical compression in bulk solids. The flow that appears below 
$600$ mK is not thermally activated, since its rate increases as the temperature decreases.
The flow rate is not proportional to the pressure difference across the solid but rather 
continues at a high rate and then abruptly stops when the final pressure is reached after 
a few minutes. This behavior is more typical of superflow than of a viscous liquid or 
conventional plastic flow.

In the recent work by Shin and Chan \cite{Shin_190}, it is discovered that blocking at a 
low temperature $T_d$ of mass flow is a temporary effect. When decreasing temperature,  
the flow commences below $0.6$ K and increases in magnitude, but then shuts off abruptly 
below a temperature $T_d$ near $0.1$ K depending on the concentration of $^3$He impurities.
The blocking temperature $T_d$ is found to increase with the concentration of $^3$He 
impurities at the few parts per million level. In the measurements \cite{Shin_190} on 
$2.5$ mm thick solid samples it is found that the mass flow rate reduction and extinction 
near $0.1$ K happens only when the concentration of the helium gas, $x_3$, used to prepare 
the sample, exceeds respectively $3.5\times 10^{-4}$ and $2\times 10^{-3}$. After the 
extinction, the mass flow shows a gradual but complete recovery with a characteristic 
time of many hours. 

There is growing and substantial evidence that the microscopic origin of the arising 
flux is due to the superfluid transport along dislocations, as has been suggested by 
Shevchenko \cite{Schevchenko_16,Schevchenko_191,Fil_192}. The flux propagates along 
dislocation cores \cite{Pollet_13,Soyler_14,Kuklov_15}. The extinction of the mass flow
is due to the trapping of $^3$He atoms at the nodes or the intersections of the 
dislocation network which blocks the transport of $^4$He along the network. The slow 
recovery in the $2.5$ mm samples, observed by Shin and Chan \cite{Shin_190}, can be due 
to the migration of the trapped $^3$He atoms along the dislocation lines and drain
into the superfluid inside the porous Vycor glass.

The earlier assumptions that the flux could be due to liquid channels that can form 
where crystal boundaries meet a container walls have been ruled out by the present 
experiments. The superfluid fraction, estimated on the basis of the observed mass flux 
of the order of a few grams per year 
\cite{Ray_5,Ray_6,Ray_7,Vekhov_8,Vekhov_9,Vekhov_10,Vekhov_11,Hallock_12}, gives the 
superfluid fraction at the level of $10^{-11}$ to $10^{-10}$, under the assumption that 
the critical velocity is about $10^4 {\rm cm}/ {\rm s}$.

It is important to emphasize that the superflow through the solid in the flow experiments 
\cite{Ray_5,Ray_6,Ray_7,Vekhov_8,Vekhov_9,Vekhov_10,Vekhov_11,Hallock_12,Hallock_185} 
has always been observed together with the giant isochoric compressibility regime, or 
the so-called syringe regime, when atoms of $^4$He from liquid are injected into the solid. 
In flow experiments one looks at the system response when pressure or temperature is 
changed in one reservoir with liquid Helium by monitoring pressure or temperature in the
second reservoir. The syringe regime is used for studying the isochoric compressibility, 
that is, the variation of the density of atoms injected into a constant-volume solid in 
response to varying the outside chemical potential. This compressibility is found to be 
comparable to that of liquid $^4$He, despite the fact that in an ideal solid it is 
essentially zero. In the syringe regime crystal growth is possible by the mechanism of 
dislocation superclimb when edge dislocations add atoms to atomic layers provided there 
is superfluid atomic transport along their cores
\cite{Soyler_14,Kuklov_15,Kuklov_17,Kuklov_193,Kuklov_194}.  

The Fermi isotope of $^3$He, in which quantum effects are even more important but 
superfluidity is not expected, has also been studied \cite{Cheng_195}. Flow measurements 
on high purity bcc $^3$He provide the possibility of a direct comparison to a solid $^4$He,
using the same cell in which a superfluidlike response in hcp $^4$He, when pressure 
differences were applied, was observed. In bcc $^3$He, the mass flow also exists, but with
rather different properties. Near melting, the flow is thermally activated, but it 
decreases monotonically with temperature. The flow rates in the solid are essentially 
constant below $100$ mK. The very different behaviors of solid $^3$He and $^4$He support 
the interpretation of superflow in $^4$He. Although superflow is not possible in $^3$He 
the temperature-independent flow below $100$ mK indicates that the flow in this regime 
also has a quantum origin, being probably due to the motion of defects and dislocations 
via thermal activation and tunneling mechanisms. It is possible also to mention the 
crowdion mechanism \cite{Pushkarov_91} when a group of atoms simultaneously slides along 
a crystallographic axis as one quantum object.

\section{Monte Carlo simulations}

Solid $^4$He has been intensively studied by means of Monte Carlo simulations. The fact 
that a perfect $^4$He crystal does not support superfluid behavior was confirmed in path 
integral Monte Carlo simulations \cite{Ceperley_18,Bernu_19,Clark_20}. For superfluid 
effects to arise, some kind of disorder is required, e.g., a layered structure or 
glass-like disorder \cite{Khairallah_21,Boninsegni_22}. We recall that throughout the paper 
we keep in mind self-organized solids, but not systems in external fields preimposing 
spatial periodicity upon constituents. Systems in external periodic fields certainly can 
be superfluid \cite{Cazorla_23}, which, however, are not real solids.

Vacancies, that in first papers were suspected as being able to induce superfluidity, 
turned out to be attractive producing phase separation instead of forming a superfluid 
fraction \cite{Boninsegni_24,Ma_25}. But superfluidity can arise at grain boundaries 
\cite{Pollet_26} and in the core of screw dislocations in $^4$He 
\cite{Soyler_14,Boninsegni_27}. Monte Carlo simulations confirmed that there is no 
superfluidity in perfect crystals, whether three-dimensional or two-dimensional 
\cite{Cazorla_28,Boninsegni_29,Rota_30,Rota_31}. The most widely accepted point of view 
is that superfluid flow at low temperatures in $^4$He occurs along dislocations forming 
a net \cite{Schevchenko_16,Schevchenko_191,Fil_192,Kuklov_17,Kuklov_194}.

\section{Solids with regions of disorder}

As is emphasized above, some kind of disorder is necessary for the occurrence of 
superfluidity in solids. For example, disorder can be caused by a net of dislocations,
with regions of disorder along these dislocations. The distribution of the latter in 
space usually is random. The description of such a solid with randomly distributed 
regions of disorder consists of two principal problems. One problem is the description 
of separate dislocations and of causes inducing superfluidity along them, which is 
discussed above. The other challenge is the statistical theory of a solid sample as a 
whole, when this sample contains randomly distributed regions of disorder inside it. 
The second problem, that is considered in the present section, aims at answering the 
question: how to describe the properties of such a sample on average, when observations 
are done over the whole sample?

In order to develop a statistical theory characterizing the sample with random regions 
of disorder, it is possible to resort to the theory of heterophase systems     
\cite{Yukalov_32,Yukalov_33,Yukalov_34,Yukalov_35,Yukalov_36,Yukalov_37,Yukalov_38}. 
The main idea of the theory is as follows. If we are interested in average properties of
a system with randomly distributed regions exhibiting a phase that is different from the
surrounding state, then it is reasonable to average over the random-region locations. 
Then we obtain a picture of a heterophase system with renormalized characteristics taking
into account the existence of two different phases.  

One should not confuse the effective heterophase system, resulting from the spatial 
averaging over mesoscopic regions with different symmetry properties, with a Gibbs 
mixture of two coexisting macroscopic phases. There are several principal differences 
between these two cases.

(i) The Gibbs mixture occurs only at phase transitions of first order, while the effective
system with nanosize inclusions of another phase can exist in a wide range of thermodynamic 
parameters. There are numerous examples of such systems. Thus, high temperature 
superconductors very often, if not always, are the composition of superconducting and 
normal regions in all range of their existence 
\cite{Phillips_195,Benedek_196,Sigmund_197,Kivelson_198,Yukalov_199,Bianconi_200}.       
The other common example is given by ferroelectrics with random regions of paraelectric 
phase existing in a wide range below the critical temperature, because of which they 
are called precursors or heterophase fluctuations
\cite{Brookeman_201,Rigamonti_202,Gordon_203,Gordon_204}. A number of other examples 
can be found in the review articles \cite{Yukalov_39,Yukalov_40,Yukalov_41}. Solid $^4$He 
with random regions of disorder along dislocations, as in the example considered above,
also exists in a wide region of temperature and pressure.  

(ii) The Gibbs mixture consists of macroscopic volumes of different phases, while the 
regions of a competing phase forming heterophase fluctuations inside the system are 
mesoscopic. These fluctuations correspond to the appearance inside a crystal of regions 
of local disorder with liquid-like properties. These fluctuations are called mesoscopic 
since their typical size $l_f$ is between the microscopic nearest-neighbor distance $a$ 
and the macroscopic system size $L$, so that $a \ll l_f \ll L$. Often, their lifetime 
$t_f$ also is mesoscopic, being between the microscopic local equilibrium time $t_{loc}$ 
and the macroscopic observation time $t_{obs}$, $t_{loc} \ll t_f \ll t_{obs}$.
In the case of a solid composed of Bose particles, mesoscopic regions of disorder, 
stretched along dislocations, can house Bose-Einstein condensate, hence, can become 
superfluid \cite{Yukalov_36}. 

(iii) The Gibbs mixture is composed of different phases occupying well defined fixed 
macroscopic volumes needing no averaging, while the mesoscopic regions of disorder are 
randomly located in space, which requires averaging over spatial configurations. From 
the mathematical point of view, the difference between the Gibbs mixture and the averaged 
effective system is evident. The Hamiltonian of the Gibbs mixture is a linear combination
of the phase Hamiltonians, while the renormalized Hamiltonian of the effective averaged 
system is not a linear combination of the terms related to different phases. 

(iv) Coexisting Gibbs macroscopic phases do not influence each other, except a thin 
layer between them. So that in the thermodynamic limit macroscopic phases are practically 
independent from each other. However, in a system with randomly distributed mesoscopic 
regions of disorder, different phases strongly influence each other. The effective 
renormalized Hamiltonian comes from the procedure of averaging over random locations of 
disordered regions. As a result of this renormalization the effective Hamiltonian acquires 
a nonlinear dependence on phase probabilities that are defined by the minimization of 
the thermodynamic potential of the whole system containing all phases. Therefore the phase
probabilities essentially depend on the properties of all phases.  

Statistical theory of a system, being from one side crystalline, and at the same time 
containing randomly distributed liquid-like regions, is based on the averaging over 
the local random configurations \cite{Yukalov_39,Yukalov_40,Yukalov_41}. This averaging,
whose mathematical details can be found in Refs. \cite{Yukalov_36,Yukalov_39}, yields 
the effective Hamiltonian $\widetilde H = H_{sol} \bigoplus H_{liq}$, being the direct 
sum of two replicas, one characterizing crystalline solid state, the other, disordered 
liquid-like state. The Hamiltonian is defined on the tensor-product space
$\widetilde \cH = \cH_{sol} \bigotimes \cH_{liq}$, where the first factor is the Hilbert 
space corresponding to the solid state and the second factor describes the liquid-like 
state.  

The number of particles in the solid phase is
$$
N_{sol} = \lgl \hat N_{sol} \rgl \; , \qquad
\hat N_{sol} = w_{sol} \int \psi_{sol}^\dgr(\br) \psi_{sol}(\br) \; d\br \; ,
$$
with $w_{sol}$ being the probability of the solid state.

In the presence of Bose-Einstein condensate, the liquid-like state has to be described 
in a self-consistent way, preserving the conditions of condensate existence and its 
stability \cite{Yukalov_42,Yukalov_43,Yukalov_44}. The field operator of the liquid-like 
state can be represented be means of the Bogolubov shift 
$\psi_{liq}(\br) = \eta(\br) + \psi_1(\br)$, with $\eta$ being the condensate function 
and $\psi_1$, the operator of uncondensed particles. The condensate function plays the 
role of the order parameter describing a system with global gauge symmetry breaking,
$\eta(\br) = \lgl \; \psi_{liq}(\br) \; \rgl$, which implies the statistical average
$\lgl \; \psi_1(\br) \; \rgl \; = 0$. The condensate function and the operator of 
uncondensed particles are mutually orthogonal.
 
The number of particles in the liquid-like state $N_{liq} = N_0 + N_1$ is the sum of 
the number of condensed particles 
$$
N_0 = w_{liq} \int |\; \eta(\br) \; |^2 \; d\br 
$$
and the number of uncondensed particles
$$
N_1 = \lgl \hat N_1 \rgl \; , \qquad
\hat N_1 = w_{liq} \int \psi_1^\dgr(\br) \psi_1(\br) \; d\br \;  ,
$$
where $w_{liq}$ is the probability of the liquid-like state.

The numbers of particles define the corresponding fractions: the solid-state fraction
$n_{sol} \equiv N_{sol}/N$, condensate fraction $n_0 \equiv N_0/N$, and the fraction of 
liquid-like uncondensed particles $n_1 \equiv N_1/N$, where the total number of particles
is $N = N_{sol} + N_{liq} = N_{sol} + N_0 + N_1$. The system chemical potential is 
$\mu = \mu_0 n_0 + \mu_1 n_1$. The phase probabilities $w_{sol}$ and $w_{liq}$ are found 
from the minimization of the grand potential of the whole system under the normalization 
condition $w_{sol} + w_{liq} = 1$.

There exists a narrow region of the system parameters, where there can arise local
Bose-Einstein condensation, hence local superfluidity \cite{Yukalov_36}. If the random 
superfluid regions form a connected net, there can arise a superfluid flow through the 
crystalline sample.

\section{Periodic droplet structures}

The ground state of a system strongly depends on its parameters and the type of particle 
interactions. Systems with dipolar interactions have recently become an object of intensive 
studies \cite{Griesmaier_45,Baranov_46,Baranov_47,Gadway_48,Kurn_49,Ueda_50,Yukalov_51}. 
Theoretical works hint on the possible existence of periodic ground-state structures with
superfluid properties for systems with dipolar interactions \cite{Lu_52}, as well as for 
some soft-core potentials \cite{Boninsegni_53} and for Rydberg-dressed atoms 
\cite{Henkel_54,Cinti_55}. The phase transition to the periodic state is usually of first 
order, although there are theoretical speculations on the existence of a critical point,
where the transition could become of second order \cite{Zhang_56}. 

Theoretical estimates predict a ground state phase diagram with three distinct regimes: 
a regular Bose-Einstein condensate, an incoherent array of droplets, and a coherent array 
of quantum droplets. The coherent droplets are connected by a background condensate, which 
leads to a phase coherence throughout the whole system. 
 
Quantum Monte Carlo simulations for dipolar systems with aligned dipolar moments have 
shown \cite{Cinti_57,Kora_58} that at low temperature there can develop triangular
periodic structures composed of filaments or droplets containing many coherent atoms 
possessing superfluid properties. Between different droplets or across different filaments 
there can arise quantum-mechanical particle exchanges allowing for the global phase 
coherence and a superfluid response. This concerns three-dimensional systems, while for 
two-dimensional dipolar systems there are indications \cite{Cinti_59} that there is no 
phase coherence among stripes, hence no global superfluid properties. 

Monte Carlo simulations also show that similar effects exist in three-dimensional Bose 
systems of Rydberg dressed atoms \cite{Seydi_60}, where there appear periodic droplet 
structures with phase coherence between droplets. A first-order quantum phase transition
from a homogeneous superfluid phase to the periodic droplet phase with face-centered 
cubic symmetry is predicted. 

In experiments, dipolar quantum droplets arranged in periodic arrays have been observed 
in trapped dipolar Bose-Einstein condensates of $^{162}$Dy, $^{164}$Dy and $^{166}$Er 
\cite{Bottcher_61,Chomaz_62,Tanzi_63,Guo_64}. The existence of periodic droplet structures 
of quantum gases with superfluid properties is confirmed by the observation of the 
low-energy Goldstone mode \cite{Tanzi_63,Guo_64}. 

The dipolar periodic structures are not absolutely stable - their lifetime is limited 
by fast inelastic losses caused by three-body collisions. Thus the droplet structure of
$^{164}$Dy survives for $0.1$ s and of $^{166}$Er, for $0.01$ s \cite{Chomaz_62}. 

Other conditions for the appearance of droplet periodic structures include the use of a 
homogeneous external magnetic field controlling the value of the scattering length
by means of the Feshbach resonance and polarizing atomic magnetic dipoles in one 
direction, which is necessary for realizing collective excitations with a roton minimum 
approaching zero. 

Under these conditions (Bose-Einstein condensation, polarization of dipoles, and tuned 
sufficiently strong dipolar interactions), the atomic system exhibits simultaneous 
breaking of global gauge symmetry and of spatial uniform symmetry, thus, forming a kind 
of a superfluid crystal.

\section{Conclusion}

One often describes a superfluid solid as a system, where spatial translational symmetry
is spontaneously broken simultaneously with the spontaneous breaking of global gauge 
symmetry. This definition correctly excludes from the sought objects those where the 
spatial symmetry is broken not spontaneously but by external fields, like in optical 
lattices, where the spatial periodicity is superimposed by laser beams. 

However this definition incorrectly excludes nonperiodic solids. Amorphous solids also 
are solids, not less than periodic crystals. Moreover, as follows from Monte Carlo studies, 
superfluidity, say in solid helium, is necessarily accompanied by the occurrence of 
defects disturbing the ideal lattice periodicity. A crystal with defects is only 
approximately periodic. In that sense, the strict periodicity is not necessary for 
superfluidity in solids, and even contradictory to it. Vise versa, these studies prove 
that for the occurrence of superfluidity in solids some breaking of periodicity is 
required.

Spontaneous breaking of global gauge symmetry is usually the sign for the existence
of superfluidity. But the breaking is not necessary: superfluidity can arise without 
this symmetry breaking, e.g. in two-dimensional systems. In general, spontaneous 
breaking of global gauge symmetry is neither necessary nor sufficient for the existence
of superfluidity \cite{Yukalov_42}. 

Mass flux experiments with solid $^4$He indicate onto the occurrence of superfluid flow 
through the solid sample. This is a bulk effect, since the mass flow along container walls 
has been ruled out by the present experiments. There is growing and substantial evidence 
that mass flow along dislocation cores is responsible for the flux. The occurrence of the 
flux is accompanied by the so-called syringe effect when the density of a solid can be 
enhanced by mass injection.  
 
Droplet dipolar crystals serve as another example of periodic systems enjoying superfluid 
properties. In these crystals, lattice spacing is self-organized, depending on the 
system parameters. One may say that it is not completely self-organized, since its 
existence requires a spacial tuning of atomic interactions, low temperature, 
and polarization of magnetic dipoles. However, actually, any crystal requires for its 
existence some external conditions, such as temperature and pressure. One may say that 
the droplet dipolar crystal is not absolutely stable, since its lifetime is limited by 
the decay time caused by three-body collisions. However, this lifetime, although short, 
anyway, is much longer than the local equilibration time. So, the system can be treated 
at least as metastable.

\section*{acknowledgments}

The author is grateful to the referees for constructive comments helping to improve the
paper. Discussions with E.P. Yukalova are appreciated.

\end{document}